\documentclass[twocolumn,showpacs,preprintnumbers,superscriptaddress,prl,floatfix]{revtex4-1}

\usepackage{amsmath,amssymb}
\usepackage{graphicx}

\renewcommand{\vec}    [1]    {\boldsymbol{#1}}

\begin{document}

\title{Impact of anharmonicity on sound wave velocities at extreme conditions}

\author{Olle Hellman}
\author{S. I. Simak}
\affiliation{Department of Physics, Chemistry and Biology (IFM), Link\"oping University, SE-581 83, Link\"oping, Sweden.}

\begin{abstract}

Theoretical calculations of sound-wave velocities of materials at extreme conditions are of great importance to various fields, in particular geophysics. For example, the seismic data on sound-wave propagation through the solid iron-rich Earth's inner core have been the main source for elucidating its properties and building models. As the laboratory experiments at very high temperatures and pressures are non-trivial, ab initio predictions are invaluable. The latter, however, tend to disagree with experiment. We notice that many attempts to calculate sound-wave velocities of matter at extreme conditions in the framework of quantum-mechanics based methods have not been taking into account the effect of anharmonic atomic vibrations. We show how anharmonic effects can be incorporated into ab initio calculations and demonstrate that in particular they might be non-negligible for iron in Earth's core. Therefore, we open an avenue to reconcile experiment and ab initio theory.
\end{abstract}

\maketitle


Sound-wave velocities of materials at extreme conditions are important parameters for building models in different physical areas, in particular in geophysics.
For example, modeling of Earth's inner core has to rely on sound-velocities from seismic data.  They, together with other cosmochemical and geochemical data, indicate that the inner core should be made out mostly of iron with approximately 10 at. \% of Ni and some light alloying elements.\cite{Geophysical1952} The choice and amount of the light elements is a rather controversial issue.\cite{Nguyen2004,Vocadlo2003}

The structural state of iron at Earth's core conditions (temperature around 6000 K and pressure around 360 GPa) is also a matter of debate. For a long time it has widely been accepted that the hexagonal close-packed  (hcp) phase of iron is the most likely  stable phase in the inner core (see Tateno et al.\cite{Tateno2010} and Tateno et al.\cite{Tateno2012} for corresponding experimental claims). Although some studies suggest that  the face-centered cubic (fcc) \cite{Vocadlo2008,Mikhaylushkin2007} or body-centered cubic (bcc) \cite{Belonoshko2003,Vocadlo2003,Dubrovinsky2007a} structures may also be stabilized at inner core conditions. The bcc structure has been strongly supported by recent theoretical works \cite{Belonoshko2017,Belonoshko2019} and has also been claimed in the latest high-pressure experiment employing a laser-heated diamond anvil cell and microstructure analysis combined with in-situ x-ray diffraction.\cite{Hrubiak2019}

A possible route to build a model of the solid inner core is to match sound velocities in laboratory experiments or computation to the observed velocities from seismic data. There have been many attempts employing such an approach in the framework of quantum-mechanics based calculations, and a variety of alloying elements and crystal structures has been considered. Remarkably, none of these attempts has led to results matching all the experimentally known peculiarities. In particular, the seismically observed velocities of the S-waves are lower than the calculated ones (by more than 10~\%). The theoretical velocities of the P-waves, however, are in good agreement with experiment.

\begin{figure*}
  \includegraphics[width=\linewidth]{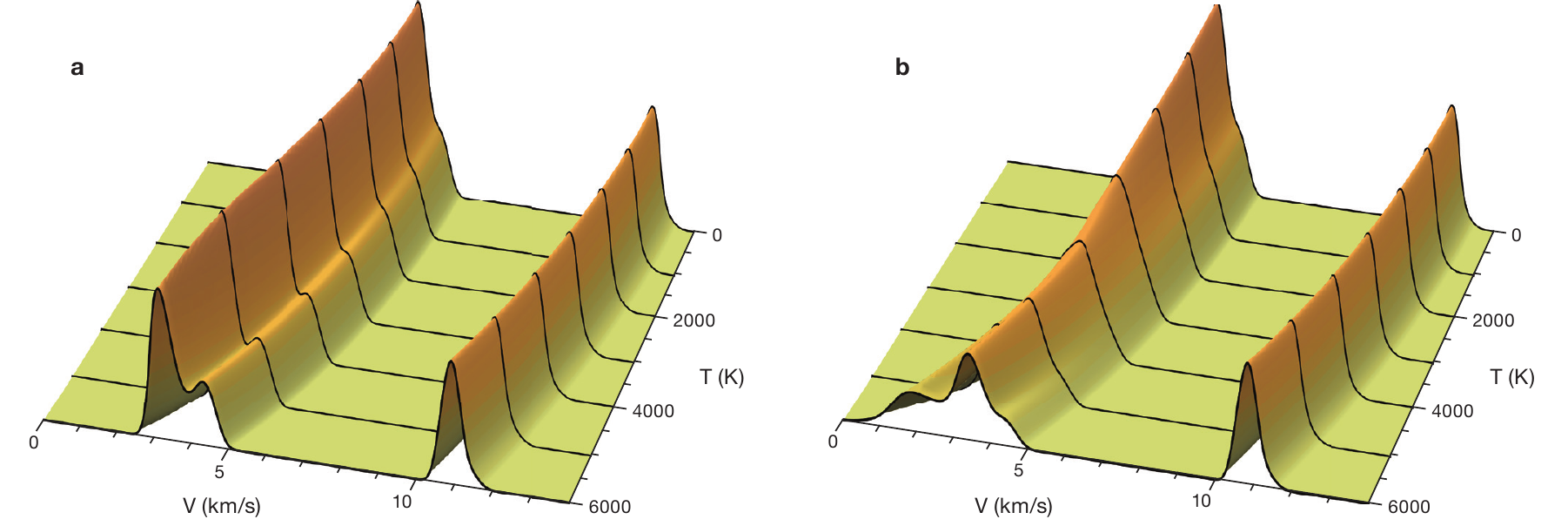}
  \caption{\label{fig:distributions} Distribution of spherically averaged sound velocities in hcp Fe at Earth's inner core pressure (360 GPa) as a function of temperature. In panel \textbf{a} the anharmonic contributions are not taken into account, and in panel \textbf{b} they are included. We identify the two regions as primary ($V_{p}$) and secondary ($V_s$) waves. One may see that the anharmonicity has little effect on $V_{p}$, but heavily influences the distribution of $V_{s}$.}
\end{figure*}

As discussed in Ref. \onlinecite{Vocadlo2007b}, if the uncertainties in the seismological values are well constrained, the difference between the observations  and theory suggests that a simple model for the inner core based on the commonly assumed phases is wrong. Different and rather complex explanations of this puzzling result have been suggested, including the partial melting of the inner core.\cite{Vocadlo2007b} It has been, however, noticed by Vo\v{c}adlo et al.\cite{Vocadlo2007b} that the disagreement between theory and experiment could in principle be accounted for by anelasticity. The simple estimates in Ref. \onlinecite{Vocadlo2007b} suggest that the effect of anelasticity should not overcome 1.5~\% and therefore has to be ruled out with the reservation that anelasticity is a very complex issue that requires material data at the conditions of Earth's inner core in order to draw irrefutable conclusions.

In this letter we show that in contrast to previous assumptions and simple models the effect of anelasticity due to anharmonic lattice vibrations may have dramatic impact on the sound-wave anisotropy of crystalline matter at high temperatures, in particular in iron at Earth's core conditions. This is proved via calculations involving the temperature-dependent effective potential (TDEP) method \cite{Hellman2011,Hellman2013,Hellman2013a}, which is based on \emph{ab initio} molecular dynamics simulations and provides accurate temperature-dependent properties of solids.


We start from the notion that the standard procedure for calculating the sound-wave velocity in a system like high-pressure high-temperature Fe is to obtain second-order elastic constants from molecular dynamics simulations at the proper temperature and density, and use these elastic constants in the Christoffel equations
\begin{equation}\label{eq:christoffel}
\vec{C}_{\vec{q}} \vec{\epsilon}_{\vec{q}} = \rho \omega^2_{\vec{q}} \vec{\epsilon}_{\vec{q}},
\end{equation}
where 
\begin{equation}\label{eq:christoffeldynmat}
C_{\vec{q}}^{\alpha\gamma}=\sum_{\beta\delta} c_{\alpha\beta\gamma\delta} q_\beta q_\delta.
\end{equation}
Here $\vec{q}$ is the wavevector of the acoustic wave, $c_{\alpha\beta\gamma\delta}$ elastic constants (any index $\alpha$, $\beta$, $\gamma$ or $\delta$ runs over Cartesian $x$, $y$ and $z$), and $\rho$ is the density. The sound velocities are obtained from the frequencies $\omega$ as $v=d\omega/dq$, and are classified via the polarization vectors $\vec{\epsilon}$. Eq.~\ref{eq:christoffel} is indeed valid, but formally only for a perfectly harmonic crystal in the long wavelength limit. In the general case, the sound waves may be perturbed, diffused and eventually changed by anharmonic lattice vibrations. In other words, the conventional approach assumes a perfect principle of superposition: any sound wave is unaffected by the intrinsic vibrations already present in the crystal -- something that is generally not true. At ambient conditions such effects are usually ignored in all but some pathological cases.\cite{Polturak2003} However, at Earth's core conditions ignoring the effects of anharmonicity can become a dangerously crude approximation.

Let us consider how anharmonicity may be accurately taken into account. One starts with the general form of the Hamiltonian describing lattice dynamics\cite{Born1998}
\begin{equation}\label{eq:b}
\begin{split}
\hat{H}= & \sum_i \frac{\vec{p}_i^2}{2m_i} + 
\sum_{ij\alpha\beta} \Phi_{ij}^{\alpha\beta}u_i^\alpha u_j^\beta + \\
& \sum_{ijk\alpha\beta\gamma} \Phi_{ijk}^{\alpha\beta\gamma} u_i^\alpha u_j^\beta u_k^\gamma + \ldots \,,
\end{split}
\end{equation}
where $p$ is the momentum, $m$ the mass, and $u_i$ the displacement of atom $i$. $\Phi$ designate interatomic force constants of increasing order. When this expansion is truncated at the second order terms, the harmonic approximation is obtained, where in the long wavelength limit the Christoffel equations (Eq.~\ref{eq:christoffel}) can be derived.

\begin{figure}
\includegraphics[width=\linewidth]{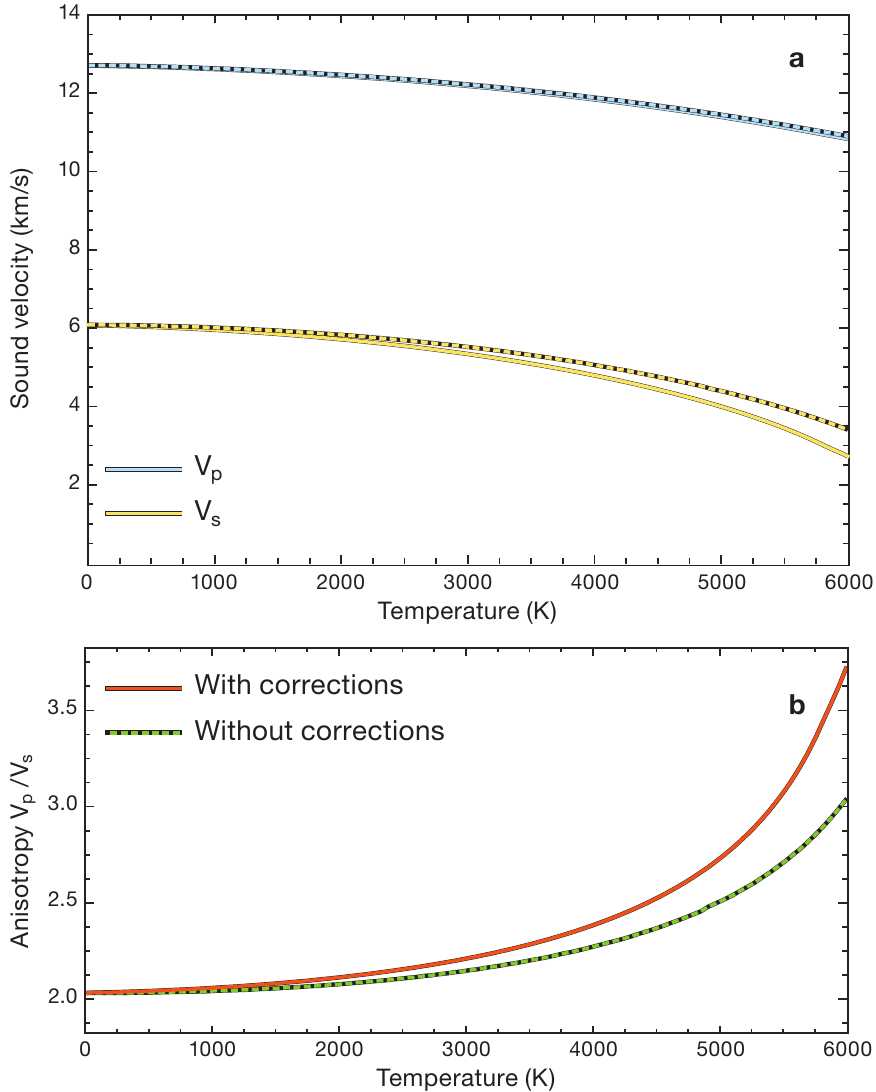}
\caption{\label{fig:soundvel_and_anisotropy}
Panel \textbf{a} shows the sound velocity in hcp Fe at Earth's inner core pressure (360 GPa) as a function of temperature. The dashed lines correspond to values calculated from panel \textbf{a} in Fig. \ref{fig:distributions}, i.e.  from the elastic constants uncorrected for anharmonic contribution.  The solid lines use the same temperature dependent elastic constants with added anharmonic corrections. The anharmonicity almost exclusively changes $V_s$, leaving $V_p$ basically unaffected. Panel \textbf{b} shows the sound velocity anisotropy as a function of temperature, indicating a strong impact of anharmonicity.}
\end{figure}

Let us now introduce strain parameters $\eta$ and write the displacement $u_i$ of an atom $i$ from its equilibrium position as
\begin{equation}\label{eq:a}
\vec{u}_i = \vec{\eta}\vec{r}_i
\end{equation}
where $\vec{r}_i$ locates atom $i$. Inserting Eq. \ref{eq:a} into Eq. \ref{eq:b} and collecting terms in powers of $\eta$ gives (to first order)
\begin{equation}
\hat{H} = 
\underbrace{
\sum_i \frac{\vec{p}_i^2}{2m_i} + 
\sum_{ij\alpha\beta} \Phi_{ij}^{\alpha\beta} u_i^\alpha u_j^\beta}
_{\hat{H}_0} + \underbrace{
\sum_{ij\alpha\beta\gamma\delta} \Phi^{\alpha \beta, \gamma \delta}_{ij}\eta_{\gamma\delta} u_i^\alpha u_j^\beta}
_{\hat{H}'}
\end{equation}
where
\begin{equation}
\Phi^{\alpha \beta, \gamma \delta}_{ij}=\sum_k \Phi_{ijk}^{\alpha \beta \gamma} r_k^\delta
\end{equation}
Here $\Phi_{ijk}^{\alpha \beta \gamma}$ are the third order interatomic force constants. $\hat{H}'$ results in a correction to the ideal harmonic waves from a strain, such as an acoustic wave. This correction expresses the change in second order force constants upon a strain. The correction is nontrivial and is best treated using perturbation theory. First, one expresses the reciprocal-space matrix elements corresponding to a strain perturbation:\cite{Cowley1967}
\begin{equation}
V^{\alpha\beta}_{\vec{q}s\vec{q}'s'}=\frac{\hbar}{4}
\sum_{ij,\gamma\delta} \frac{
\vec{\epsilon}_{\vec{q}s}^{i\gamma} \vec{\epsilon}_{\vec{q}'s'}^{j\delta}
}{
\sqrt{\omega_{\vec{q}s} \omega_{\vec{q}'s' m_i m_j}} 
}
\Phi^{\alpha \beta, \gamma \delta}_{ij}
e^{i\vec{q}\vec{r}_i+i\vec{q}'\vec{r}_j}
\end{equation}
Here $\omega_{\vec{q}s}$ and $\vec{\epsilon}_{\vec{q}s}$ stand for phonon frequencies and polarization vectors at reciprocal momentum $\vec{q}$ for the phonon branch $s$, respectively. As can be shown (see Ref. \onlinecite{Cowley1967} and Ref. \onlinecite{RACowley1968}), the $C_{\vec{q}}$ tensor in Eq. \eqref{eq:christoffeldynmat} should now contain 3 additional terms due to anharmonicity
\begin{align}
\label{eq:Cdelta1}
\Delta^1 c_{\alpha \beta, \gamma \delta} &=\frac{8}{V\hbar}\sum_{\vec{q}}\sum_{ss'}
V^{\alpha\beta}_{\vec{q}s\bar{\vec{q}}s'}
V^{\gamma\delta}_{\vec{q}s\bar{\vec{q}}s'}
\frac{\omega_{\vec{q}s}(2n_{\bar{\vec{q}}s'}+1)}
{\omega_{\bar{\vec{q}}s'}^2-\omega_{\vec{q}s}^2}
\\
\label{eq:Cdelta2}
\Delta^2 c_{\alpha \beta, \gamma \delta} &=
-\frac{2}{V\hbar}\sum_{\vec{q}}\sum_{s}
V^{\alpha\beta}_{\vec{q}s\bar{\vec{q}}s}
V^{\gamma\delta}_{\vec{q}s\bar{\vec{q}}s}
\frac{2n_{{\vec{q}}s}+1}
{\omega_{\vec{q}s}}
\\
\intertext{and the final term that is either zero for adiabatic elastic constants or in the isothermal case is equal to}
\label{eq:Cdelta3}
\Delta^3 c_{\alpha \beta, \gamma \delta} &=
\frac{4\beta}{V}\sum_{\vec{q}}\sum_{s}
V^{\alpha\beta}_{\vec{q}s\bar{\vec{q}}s}
V^{\gamma\delta}_{\vec{q}s\bar{\vec{q}}s}
(n_{\vec{q}s} + 1) n_{\vec{q}s}
\end{align}
In Eqs. \ref{eq:Cdelta1}, \ref{eq:Cdelta2} and \ref{eq:Cdelta3} $n_{\vec{q}s}$ is the occupation number of the phonon with reciprocal momentum $\vec{q}$ for the phonon branch $s$ and therefore explicitly depends on temperature. We use $\bar{\vec{q}}$ to denote $-\vec{q}$.

\begin{figure*}
\includegraphics[width=\linewidth]{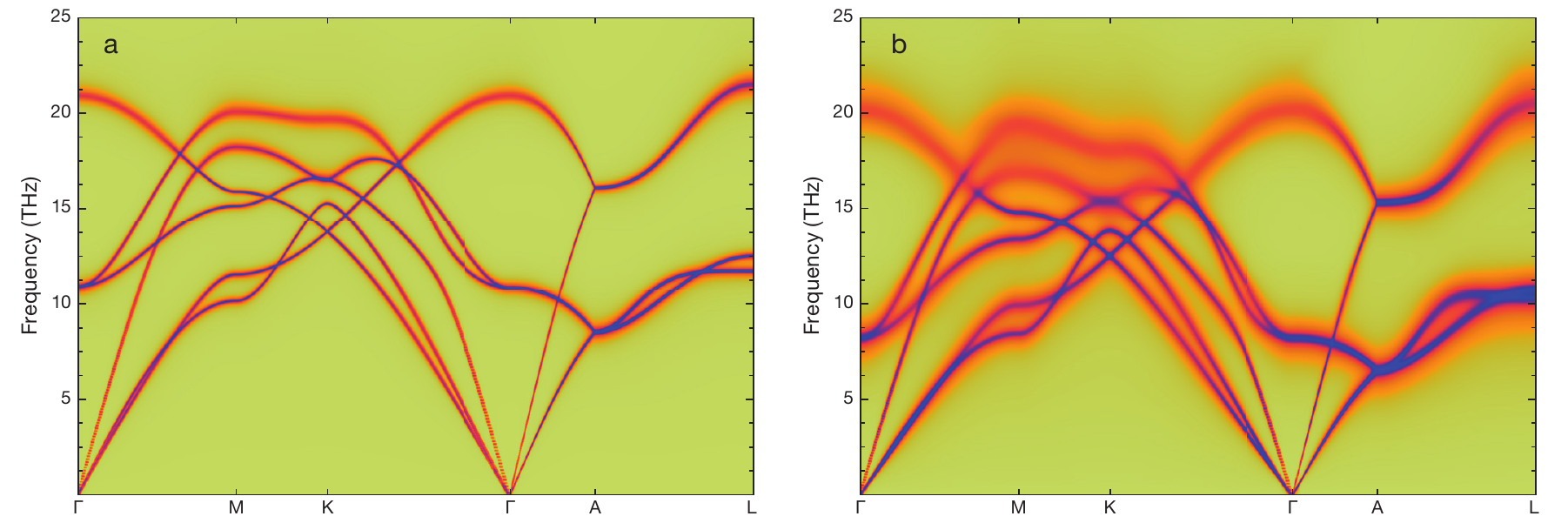}
\caption{\label{fig:two_lineshapes}
The broadened phonon spectrum of hcp Fe at Earth's inner core pressure (360 GPa). Panel \textbf{a} is at 2000K, and panel \textbf{b} at 6000K. One may see that the broadening of the low-lying transverse acoustic modes is more affected by temperature than the broadening of the higher-lying longitudinal acoustic modes. The slopes of the former in $\Gamma$ contribute most to the S-wave sound velocity. Accordingly, the observed broadening leads to the decrease of the S-wave sound velocity with temperature bringing in agreement theory and experiment.}
\end{figure*}

There is a feature of Eqs. \ref{eq:Cdelta1},\ref{eq:Cdelta2} and \ref{eq:Cdelta3}, which plays a crucial role: the corrections are highly anisotropic. Therefore, even if the corrections may be rather small, they can have a significant impact on the sound-wave anisotropy. We stress that these are not corrections to the macroscopic elastic constants, but corrections to the sound-wave velocities.

We notice that the theoretical formalism above has been known for many decades (see Refs. \onlinecite{Cowley1967} and \onlinecite{RACowley1968} for the full derivation of general results). However, its practical usage has been hindered by the inability to calculate accurate interatomic force constants up to the third order. The temperature dependent effective potential technique (TDEP)\cite{Hellman2013a,Hellman2011,Hellman2013} allows one to achieve this goal in a rigorous and efficient way.


Actual calculations have been done for hcp Fe, which is expected to be the least anharmonic phase out of possible candidates for Earth's core conditions. That is if any effect on sound velocities is observed for hcp Fe, it should be expected to be even larger for other iron phases. We also notice that the hcp case is not complicated by collective self-diffusion events predicted for bcc Fe.\cite{Belonoshko2017}  The calculations have been performed at a pressure of 360 GPa corresponding to the one in Earth's inner core\cite{Dziewonski1981} in the temperature range from 0 to 6000 K. \emph{Ab initio} molecular dynamics simulation has been run on a 96 atom $4 \times 4 \times 3$ atom supercell in the framework of Density Functional Theory (DFT) in its generalized gradient approximation form PBE\cite{Perdew1996} and Projector-Augmented Wave method\cite{Blochl1994a} as implemented in Vienna Ab initio Simulation Package (VASP).\cite{Kresse1999,Kresse1996c,Kresse1996,Kresse1993b} The plane-wave cut-off was set to 500 eV. We used a 1 fs time-step and ran the simulations using the $2 \times 2 \times 2$ k-points for the BZ integration for 5000 time-steps after equilibration. Interatomic force constants up to the third order were calculated as described in Refs. \onlinecite{Hellman2013a,Hellman2013}, and the anharmonic corrections to the sound-wave velocities were obtained from Eqs. \ref{eq:Cdelta1},\ref{eq:Cdelta2} and \ref{eq:Cdelta3} with a $31 \times 31 \times 31$ $q$-point grid for the Brillouin zone integrations. For further details on determining the elastic constants see the supplementary information.


The calculated temperature dependence of the longitudinal and transverse sound-wave velocities with and without the total anharmonic corrections is presented in Fig. \ref{fig:soundvel_and_anisotropy}. The uncorrected elastic constants were calculated in the same way as in Ref. \onlinecite{Vocadlo2009} (see the supplementary information for details) and the anharmonic corrections were added on top of them.  The averaging was done by spherical sampling and constructing histograms, where the center of mass of the top region was identified as $V_p$ and lower region as $V_s$. These averages are presented in Fig. \ref{fig:distributions}.

As one may see (Fig. \ref{fig:distributions}, Fig. \ref{fig:soundvel_and_anisotropy}), the impact of anharmonic corrections on the longitudinal P-wave velocity is basically negligible in the whole range of temperatures. The situation with the transverse S-wave is, however, strikingly different. Though up to about 2000 K the effect is minor, it gives up to 10~\% contribution at 6000 K. This temperature-induced anisotropy is clearly seen in Fig. \ref{fig:soundvel_and_anisotropy}b, where curves correspond to the ratio between $V_p$ and $V_s$.

The physical reason behind the effect can be traced down by considering how  temperature affects the phonon life-time and accordingly broadening of the phonon dispersion relations. As one may see from Fig. 
\ref{fig:two_lineshapes}, the broadening of higher-lying longitudinal acoustic modes (the frequency goes to 0 in $\Gamma$) in $\Gamma - M$ and $\Gamma - A$ directions, i.e. those whose slopes at $\Gamma$ correspond to $c_{11}$ and $c_{33}$ elastic constants, respectively, is nearly negligible even at 6000 K. However, the broadening of the lower-lying transverse acoustic modes in the $\Gamma - M$ and $\Gamma - K$ directions, i.e. those whose slopes at $\Gamma$ correspond to $c_{66}$ and $c_{44}$ elastic constants, respectively, is visibly increased. This reveals that the lower-energy phonon modes, contributing most to the S-wave sound velocities, are more affected by the high-temperature anharmonic lattice vibrations than those contributing most to the P-wave counterparts.

Certainly, the presented results bear no claim concerning the structure of the inner core. Even though the anharmonic corrections bring the anisotropy into the observed range, some other properties of hcp Fe do not match observations.\cite{Belonoshko2019} It would be highly desirable to perform similar calculations for bcc-Fe at Earth's core conditions, which appears as a good candidate according to Refs. \onlinecite{Belonoshko2017,Belonoshko2019}. However, one needs first to find a way to treat Fe self-diffusion, which is shown in Ref.\cite{Belonoshko2017} to be a crucial part of bcc-Fe stabilization at Earth's core conditions.

What we show in this letter is that anharmonic effects on sound wave velocities cannot be ignored under high-pressure - high-temperature conditions. The effect on the sound wave anisotropy is not exclusive to pure iron in the hcp structure, it is a general feature. In particular, any comparisons with seismic data should take the possibility of anharmonicity into account. The methodology presented in this letter is readily applicable to other crystals, and opens up the possibility to revisit previously calculated data. We also note that the proposed methodology is applicable to any problem where sound wave propagation at finite temperature is of interest, e.g. ultrasonic attenuation.


The support the Swedish Government Strategic Research Area in Materials Science on Advanced Functional Materials at at Link{\"o}ping University (Faculty Grant SFO-Mat-LiU No.\ 2009-00971) is acknowledged by S.I.S.  The computations were performed on resources provided by the Swedish National Infrastructure for Computing (SNIC) at the PDC Centre for High Performance Computing (PDC-HPC) and the National Supercomputer Center (NSC).


%

\end{document}